\documentclass{article}
\usepackage[utf8]{inputenc}

\title{On the Upper Bound of the Kullback-Leibler Divergence and Cross Entropy}
\author{Min Chen$^1$ and Mateu Sbert$^2$\\~\\
$^{1}$ University of Oxford, UK; min.chen@oerc.ox.ac.uk\\
$^{2}$ University of Girona, Spain; mateu@ima.udg.edu}

\date{November 2019}

\usepackage{amsfonts}
\usepackage{amsmath}
\usepackage{amssymb}
\usepackage{stmaryrd}

\usepackage[utf8]{inputenc}
\usepackage[english]{babel}

\usepackage{graphicx}
\usepackage{color}

\begin{document}
\maketitle

\begin{abstract}
This archiving article consists of several short reports on the discussions between the two authors over the past two years at Oxford and Madrid, and their work carried out during that period on the upper bound of the Kullback-Leibler divergence and cross entropy. The work was motivated by the cost-benefit ratio proposed by Chen and Golan \cite{Chen:2016:TVCG}, and the less desirable property that the Kullback-Leibler (KL) Divergence used in the measure is unbounded. The work subsequently (i) confirmed that the KL-Divergence used in the cost-benefit ratio should exhibit a bounded property, (ii) proposed a new divergence measure, and (iii) compared this new divergence measure with a few other bounded measures.
\end{abstract}

\section{Background and Motivation}

The mathematical definitions of Kullback-Leibler (KL) Divergence and Cross Entropy do not imply an upper bound.
Consider a simple alphabet $\mathbb{Z}=\{\mathbf{z}_1, \mathbf{z}_2\}$, and two probability mass functions (PMFs) defined upon $\mathbb{Z}$: $P=\{0+\epsilon,1-\epsilon\}$ and $Q=\{1+\epsilon,0+\epsilon\}$, where $0 < \epsilon < 1$.
When $\epsilon \rightarrow 0$, we have $\mathcal{D}_{\text{KL}}(P||Q) \rightarrow \infty$ and $\mathcal{H}(P, Q) \rightarrow \infty$.

Chen and Golan proposed an information-theoretic measure for analyzing the cost-benefit of machine- and human-centrics in data intelligence workflows \cite{Chen:2016:TVCG}.
Given a process $\mathbf{P}_i$ with $\mathbb{Z}_i$ as its input alphabet and $\mathbb{Z}_{i+1}$ as its output alphabet, the cost benefit ratio of $\mathbf{P}_i$ is as follows:
\begin{equation}\label{eq:CBR}
\begin{split}
    \frac{\text{Benefit}}{\text{Cost}}
    &= \frac{\text{Alphabet Compression}-\text{Potential Distortion}}{\text{Cost}}\\
    &= \frac{\mathcal{H}(\mathbb{Z}_i) - \mathcal{H} (\mathbb{Z}_{i+1}) + \mathcal{D}_{\text{KL}}(\mathbb{Z}'_i||\mathbb{Z}_i)}{\text{Cost}}
\end{split}
\end{equation}
\noindent where $\mathbb{Z}'_i$ is an alphabet reconstructed based on $\mathbb{Z}_{i+1}$ by a reverse process $\mathbf{P}_i^{-1}$.
It has the same letters as $\mathbb{Z}_i$ but is likely to have a different PMF since in data intelligence, $\mathbf{P}_i$ will bring about information loss.   

Because the alphabet compression part, $\mathcal{H}(\mathbb{Z}_i) - \mathcal{H} (\mathbb{Z}_{i+1})$, is bounded when $||\mathbb{Z}_i||$ and $||\mathbb{Z}_{i+1}||$ are finite but the potential distortion part, $\mathcal{D}_{\text{KL}}(\mathbb{Z}'_{i}||\mathbb{Z}_{i})$, is unbounded, it is difficult to interpret the meaning of the benefit when it approaches $-\infty$.

Chen and Golan discussed this issue when they were working on Equation \ref{eq:CBR}, and agreed that the issue could be addressed by using the conventional method employed in many practical applications of information theory, i.e., by fixing an $\epsilon$ as the lower bound of probability values in a PMF, $\forall p \in P$, $\epsilon \le p \le (1-\epsilon)$.

In the analysis of the results of an empirical study, Kijmongkolchai et al. converted the values of accuracy and response time to benefit and cost in Equation \ref{eq:CBR}.
They set the upper bound as $\max \bigl( \mathcal{D}_{\text{KL}}(P||Q) \bigr) = 2\mathcal{H}_\text{max}(P)$, such that the benefit is bounded by $[-\mathcal{H}_\text{max}(P),+\mathcal{H}_\text{max}(P)]$.
However, this raises a number of questions, such as:
\begin{enumerate}
    \item Should there be an upper bound? (See Section \ref{sec:Upper}.)
    \item If (1) is true, should $\mathcal{D}_{\text{KL}}(\mathbb{Z}'_i||\mathbb{Z}_i)$ be replaced with a bounded measure? (see Section \ref{sec:New}.)
    \item If (1) is true, what is the most suitable bounded divergence? (see Section \ref{sec:Analysis}.)
\end{enumerate}

\section{The Existence of an Upper Bound}
\label{sec:Upper}
During his research visit to Oxford in July 2017, Sbert suggested to examine the upper bound based on the semantics of coding associated with the cross entropy.
On 19 July 2017, they discovered that there is an upper bound of the value $||\mathbb{Z}||-1$ of $\mathcal{H}(P, Q)$.

Let $\mathbb{Z} = \{\mathbf{z}_1, \mathbf{z}_2, \ldots, \mathbf{z}_n\}$ be an alphabet, which is associated with a PMF, $Q$, such that:
\begin{align*}
q(\mathbf{z}_n) &= \epsilon, \quad\text{(where $0 < \epsilon < 2^{-(n-1)}$}),\\
q(\mathbf{z}_{n-1}) &= (1-\epsilon)2^{-(n-1)},\\
q(\mathbf{z}_{n-2}) &= (1-\epsilon)2^{-(n-2)},\\
&\cdots\\
q(\mathbf{z}_{2}) &= (1-\epsilon)2^{-2},\\
q(\mathbf{z}_{1}) &= (1-\epsilon)2^{-1} + (1-\epsilon)2^{-(n-1)}.
\end{align*}

When we encode this alphabet using an entropy binary coding scheme, we can be assured to achieve an optimal code with the lowest average length for codewords.
One example of such a code for the above probability is:

\begin{align*}
\mathbf{z}_1&: 0\\
\mathbf{z}_2&: 10\\
\mathbf{z}_3&: 110\\
&\cdots\\
\mathbf{z}_{n-1}&: 111\ldots10 \quad\text{(with $n-2$ ``1''s and one ``0'') }\\
\mathbf{z}_n&: 111\ldots11 \quad\text{(with $n-1$ ``1''s and no ``0'') }
\end{align*}

In this way, $\mathbf{z}_n$, which has the smallest probability, will always be associated with a codeword with the maximal length of $n-1$.
Regardless whatever a PMF is defined upon $\mathbb{Z}=\{\mathbf{z}_1, \mathbf{z}_2, \ldots, \mathbf{z}_n\}$,
there is no need to code any letter $z_i \in \mathbb{Z}$ with more than $n-1$ bits.
The entropy coding scheme is designed to minimize the number of bits to be transmitted over a communication channel for sending a ``very long'' sequence of letters in the alphabet.
The phrase ``very long'' implies that the string exhibits the PMF used in the coding.
However, for any other string that does not exhibit the PMF used in the coding, there will be inefficiency.
This inefficiency is usually measured using cross entropy.
Let $P$ be the PMF of such a string, and $P$ may be different from $Q$.
The inefficiency is measured by:
\[
    \mathcal{H}(P, Q) = -\sum_{i=1}^n p_i \log_2 q_i
    = \mathcal{H}(P) + \mathcal{D}_{\text{KL}}(P||Q)
\]

Clearly, the worst case is that the letter, which was encoded using the most number of bits, $n-1$, turns out to be the most frequently used letter.
It is so frequent that all letters in the string are of this letter.
The average codeword length of this string is thus of $n-1$ bits.
Since there is no informative variation in the PMF $P$ for this very long string, i.e., $\mathcal{H}(P) = 0$, in principle, the transmission of this string (of $n-1$ bits per letter) is unnecessary.
The situation cannot be worse.
Therefore $n-1=||\mathbb{Z}||-1$ is an upper bound $\top_{\text{CE}}$ for the cross entropy.

An upper bound, $\top_{\text{KL}}$, of the KL-Divergence between $P$ and $Q$, can be derived from:
\begin{equation}\label{eq:CEandKL}
	\mathcal{D}_{\text{KL}}(P||Q) = \mathcal{H}(P, Q) - \mathcal{H}(P) \le \top_{\text{CE}} - \min_{\forall P}\bigl( \mathcal{H}(P) \bigr)
\end{equation}
\noindent In the cases where all PMFs are possible, the minimal Shannon entropy is 0.
Hence we have $\top_{\text{KL}} = \top_{CE}$.

There is a special case worth mentioning.
In practice, it is common to assume that $Q$ is a uniform PMF, i.e., $q_i = 1/n, \forall q_i \in Q$, typically because $Q$ is unknown or varies frequently.
Hence the assumption leads to a code with an average length equaling the maximum entropy $\mathcal{H}_{\text{max}} = \log_2 (n)$.
In practice, the actual code length would be $\lceil \log_2 (n) \rceil$.

Under this special (but rather common) condition, all letters in a very long string have codewords of the same length.
The worst case is that all letters in the string turn out to the same letter.
Since there is no informative variation in the PMF $P$ for this very long string, i.e., $\mathcal{H}(P) = 0$, in principle, the transmission of this string is unnecessary. 
The maximal amount of inefficiency is thus $\log_2 (n)$ or in practice $\lceil \log_2 (n) \rceil$.
This is indeed much lower than the upper bound $\top_{\text{CE}} = n-1$, justifying the assumption or use of a uniform $Q$ in many situations.


\section{A New Bounded Divergence}
\label{sec:New}
During his research visit to Madrid in July 2019, Chen discussed a possibly new measure with Sbert.
They conducted a literature study, and did not find anything similar in the literature.

Given two PMFs, $P$ and $Q$, associated with the same alphabet $\mathbb{Z}$, the new measure is:
\begin{equation} \label{eq:New}
    \mathcal{D}_{\text{new}}(P||Q) = \sum_{i=1}^n p_i \log_2 \bigl( |p_i - q_i| + 1 \bigr)
\end{equation}

\noindent
Like $\mathcal{D}_{\text{KL}}(P||Q)$, $\mathcal{D}_\text{new}(P||Q)$ is not commutative.
The cost-benefit ratio in Equation (\ref{eq:CBR}) does not require it to be commutative because it is about a reconstruction process that maps its output alphabet to its input alphabet but not vice versa.
Importantly, $\mathcal{D}_{\text{new}}(P||Q)$ is bounded.
Obviously, $\mathcal{D}_{\text{new}}(P||Q) \ge 0$, and it equals 0 if and only if $P \equiv Q$.
Because $|p_i - q_i| < 1$, we have
\[
    \mathcal{D}_{\text{new}}(P||Q) = \sum_{i=1}^n  p_i \log_2 \bigl( |p_i - q_i| + 1 \bigr)
    \le \sum_{i=1}^n p_i \log_2 (2) = 1
\]

If we replace $\mathcal{D}_{\text{KL}}(\mathbb{Z}'_i||\mathbb{Z}_i)$ with $\mathcal{D}_{\text{new}}(\mathbb{Z}'_i||\mathbb{Z}_i)$, it is necessary to scale its value with $\mathcal{H}_{\text{max}}(\mathbb{Z}_i)$.
In other words, Equation (\ref{eq:CBR}) can be rewritten as:
\begin{equation}\label{eq:NewCBR}
\begin{split}
    \frac{\text{Benefit}}{\text{Cost}}
    &= \frac{\text{Alphabet Compression}-\text{Potential Distortion}}{\text{Cost}}\\
    &= \frac{\mathcal{H}(\mathbb{Z}_i) - \mathcal{H}(\mathbb{Z}_{i+1}) + \mathcal{H}_{\text{max}} \mathcal{D}_{\text{new}}(\mathbb{Z}'_i||\mathbb{Z}_i)}{\text{Cost}}
\end{split}
\end{equation}

 This new version can be used to measure the benefit with values that can be interpreted intuitively.
 For example, consider a type of things (or situations) that have two states, good or bad.
 The actual PMF is $\{ 0.7, 0.3 \}$, that is, 70\% good and 30\% bad.
 A process $\mathbf{P}_i$ conveys a ``simplified'' binary message that these things are always good, i.e., with a PMF $\{ 1, 0 \}$.
 \begin{itemize}
     \item If the audience is totally misled to think that the things are always good, we have:
     $\text{Benefit} =  0.88 - 0 - 0.38 = 0.5$ bits.
     \item If half of the audience have the knowledge about the ground truth, the reconstructed PMF becomes $(0.85, 0.15)$, the benefit of $\mathbf{P}_i$ improves from 0.5 to: $\text{Benefit} =  0.88 - 0 - 0.20 = 0.68$ bits.
     \item If all of the audience have the knowledge about the ground truth, the information loss due to the binary message does not cause any potential distortion. The benefit is thus: $\text{Benefit} =  0.88 - 0 - 0 = 0.88$ bits.
 \end{itemize}
 
 Consider another example, where a type of things (or situations) are always good, that is, for the input, the PMF $\{ 1, 0 \}$.
 A process $\mathbf{Q}_i$ conveys a misleading message that these things are always bad, i.e., for the output, the PMF is $\{0, 1\}$.
 \begin{itemize}
     \item If the audience is totally misled to think that the things are always bad, we have:
     $\text{Benefit} =  0 - 0 - 1 = -1$ bits. Note that with the KL-divergence, the benefit would approach $-\infty$.
     \item If half of the audience notice a note of sarcasm in the message, and interpret that the things are actually good. The reconstructed PMF becomes $(0.5, 0.5)$ and the benefit is: $\text{Benefit} =  0 - 0 - 0.58 = -0.58$ bits.
     \item If all of the audience correctly interpret the message to be that the things are always good, the benefit is thus: $\text{Benefit} =  0 - 0 - 0 = 0$ bits.
 \end{itemize}
 
 A generalised version of $\mathcal{D}_{\text{new}}$ is
\begin{equation} \label{eq:NewG}
    \mathcal{D}^k_{\text{newG}}(P||Q) = \sum_{i=1}^n p_i \log_2 \bigl( |p_i - q_i|^k + 1 \bigr)
\end{equation}

A commutative version of $\mathcal{D}^k_{\text{newG}}$ is
\begin{equation} \label{eq:NewGC}
    \mathcal{D}^k_{\text{newGC}}(P||Q) = \frac{1}{2} \sum_{i=1}^n \bigl( p_i + q_i \bigr) \log_2 \bigl( |p_i - q_i|^k + 1 \bigr)
\end{equation}

\section{Comparing Several Bounded Measures}
\label{sec:Analysis}
During the meeting in Madrid in July 2019, Chen and Sbert also compared the measure with several other measures in the literature.
These include the Jensen-Shannon (JS) divergence and the Minkowaski distances.

\begin{figure}[t]
    \centering
    \includegraphics[width=.94\linewidth]{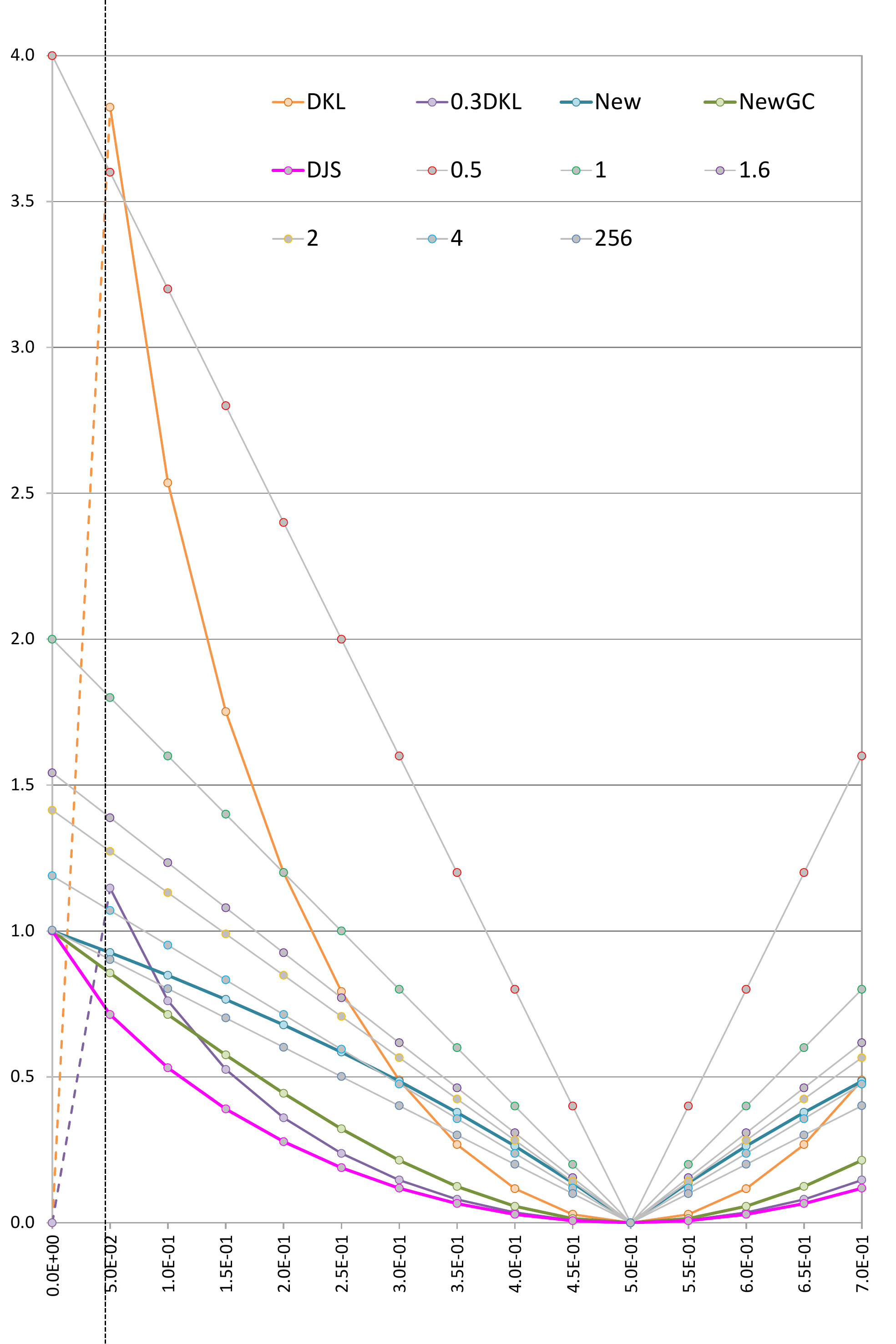}
    \caption{Comparing the measures of divergence in relation to an alphabet in the range [0.05, 0.7]. The curve segments of DKL and 0.3DKL in the range [0, 0.05] do not represent the actual shapes.}
    \label{fig:NonZero}
\end{figure}

Figure \ref{fig:NonZero} shows the measures of
\begin{itemize}
    \item the KL-divergence (as DKL),
    \item its scaled down measures (as 0.3DKL),
    \item the new divergence in Equation (\ref{eq:New}) (as New),
    \item the generalized and commutative version of the new divergence with $k=2$ (as NewGC),
    \item the JS-divergence (as DJS), and
    \item several Minkowaski distances with $k = 0.5, 1, 1.6, 2, 4, 256$.
\end{itemize}
The $x$-axis shows the value of $\epsilon$ in the range of [0.05, 0.7].
The two PMFs are set as $P=\{1-\epsilon, \epsilon\}$ and $Q=\{\epsilon, 1-\epsilon\}$.
Therefore, $P=Q$ when $\epsilon=0.5$, and their divergence increases when $\epsilon \rightarrow 0$ or $\epsilon \rightarrow 1$.

\begin{figure}[t]
    \centering
    \includegraphics[width=.94\linewidth]{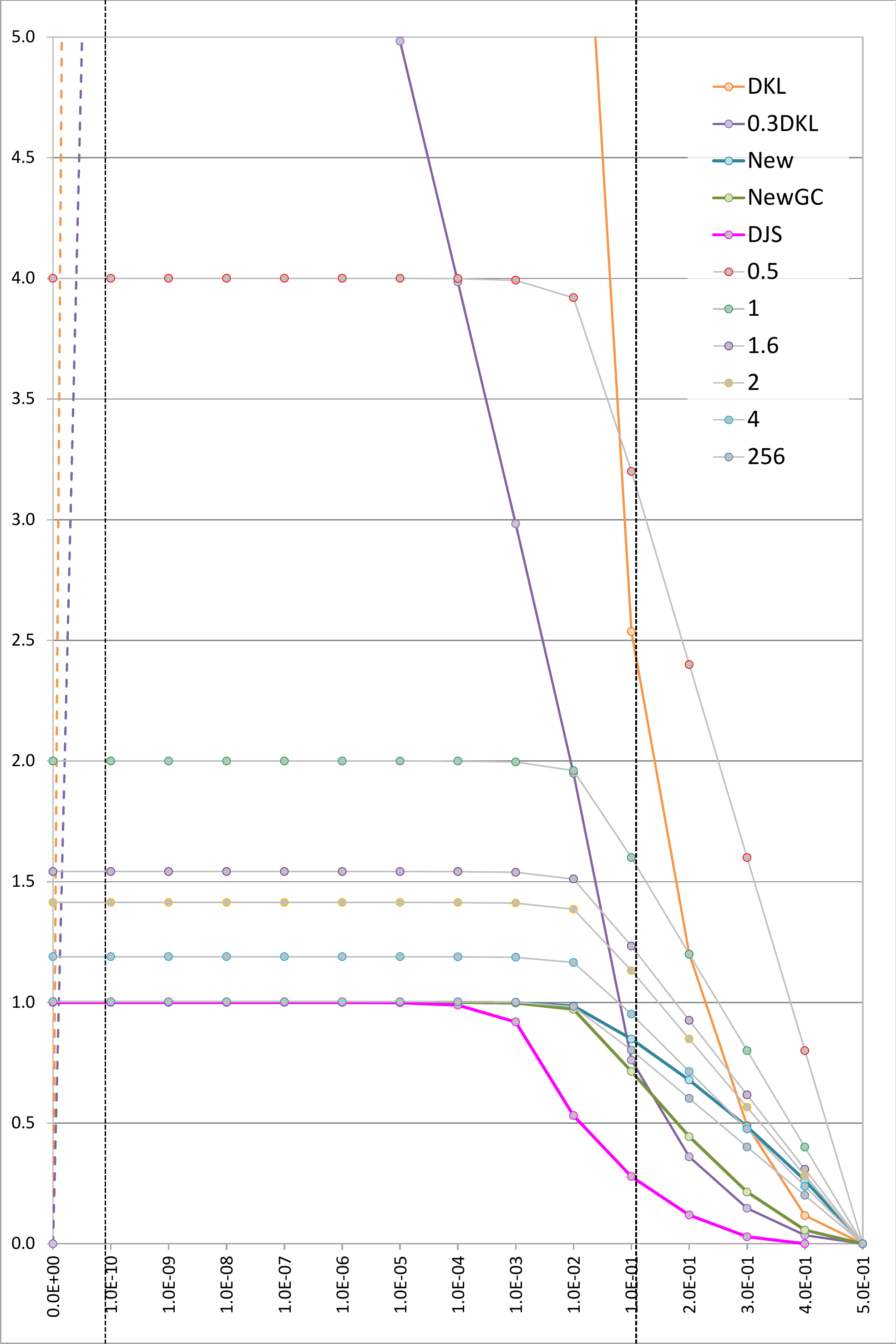}
    \caption{Comparing the measures of divergence in relation to an alphabet in the range near zero. The curve segments of DKL and 0.3DKL in the range $[0, 0.1^{10}]$ do not represent the actual shapes. The ranges $[0, 0.1^{10}]$ and $[0.1, 0.5]$ are only for references as they do not use the same logarithmic scale as in the range $[0.1^{10}, 0.1]$.}
    \label{fig:NearZero}
\end{figure}

\clearpage

We can observe that the curve of DKL quickly moves about 1.0, which is the maximal entropy of $P$ and $Q$.
When we scale the values of DKL to the one-third of its values, we can observe that the curve of 0.3DKL is similar to those of DJS and NewGC.
Meanwhile, the Minkowaski distances do not seem to capture much features of DKL.
The curve of the basic version of the new divergence in Equation (\ref{eq:New}) seems to differ from those of 0.3DKL, DJS, and NewGC.

Figure \ref{fig:NearZero} shows the same set of measures in the range near zero, that is, $\epsilon$ varies from $0.1^{10}$ to $0.1$. The ranges $[0, 0.1^{10}]$ and $[0.1, 0.5]$ are there only for references as they do not have the same logarithmic scale as that in the range $[0.1^{10}, 0.1]$.
We can observe that in $[0.1^{10}, 0.1]$ the curve of 0.3DKL also rises quickly as DKL.
This confirms that simply scaling the KL-divergence is not an adequate solution.

The curves of New and NewGC converge earlier than that of DJS.
If the curve of 0.3DKL is used as a benchmark as in Figure \ref{fig:NonZero}, the curve of NewGC is closer to 0.3DKL than that of DJS.

From Figures \ref{fig:NonZero} and \ref{fig:NearZero}, we can see that DJS (the JS-divergence) and NewGC (the commutative version of the new divergence with $k=2$) are the better options as bounded measures to replace the KL-divergence in Equation (\ref{eq:CBR}).

\begin{figure}[t]
    \centering
    \includegraphics[width=0.95\linewidth]{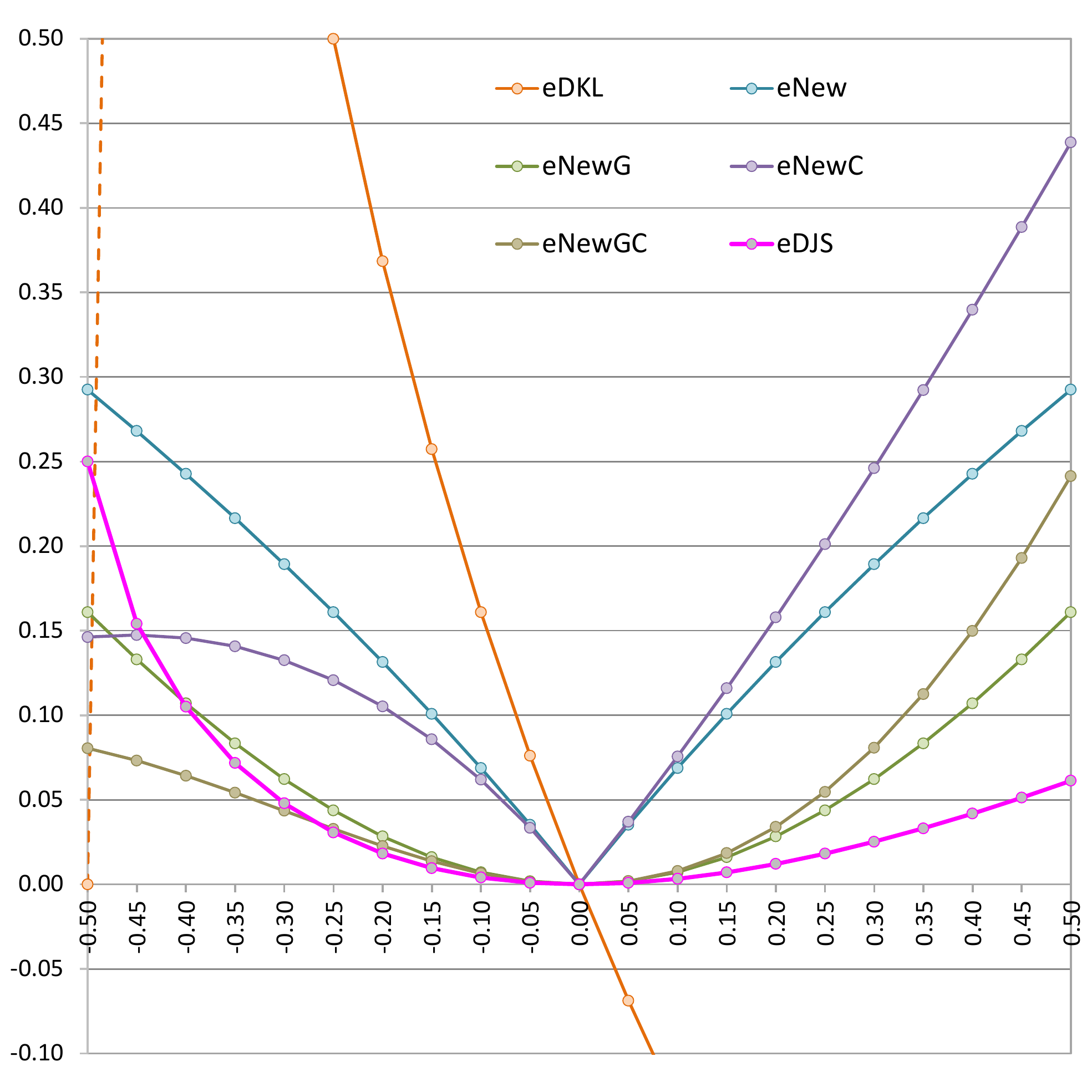}
    \caption{Comparing the measures of divergence in relation to an individual letter $z_i \in \mathbb{Z}$.
    The $y$-axis shows only one term related to $p_i$ and $q_i$ but not the whole divergence measure.}
    \label{fig:Element}
\end{figure}

Figure \ref{fig:Element} compares the measures for an individual letter $z_i \in \mathbb{Z}$.
We set $p_i = 0.5$ while varying $q_i$ from 0 to 1.
The $x$-axis shows the difference $\delta = q_i - p_i = q_i - 0.5$.
The $y$-axis shows the measures returned by:
\begin{itemize}
    \item eDKL: $p_i \log_2 \frac{p_i}{q_i}$ --- an element of the KL-divergence;
    \item eNew: $p_i \log_2 (|p_i - q_i|+1)$ --- an element of the non-commutative version of the new divergence (i.e., $k=1$);
    \item eNewG: $p_i \log_2 (|p_i - q_i|^2+1)$ --- an element of the generalized version of the new divergence with $k=2$;
    \item eNewC: $\frac{1}{2} (p_i+q_i) \log_2 (|p_i - q_i|+1)$ --- an element of the commutative version of the new divergence;
    \item eNewGC: $\frac{1}{2} (p_i+q_i) \log_2 (|p_i - q_i|^2+1)$ --- an element of the generalized and commutative version of the new divergence with $k=2$;
    \item eDJS: $\frac{1}{2} \bigl( p_i \log_2 \frac{2 p_i}{p_i + q_i} + q_i \log_2 \frac{2 q_i}{p_i + q_i} \bigr)$ --- an element of the JS-divergence.
\end{itemize}

We can make the following observations:

\begin{enumerate}
    \item The trend towards infinity and the negative values of eDKL are not intuitive to interpret semantically.
    \item The asymmetric patterns shown by eNewC, eNewGC, and eDJS indicates that these measures are influenced not only by the difference or ratio between $p_i = 0.5$ and $q_i \in [0, 1]$, but also by the combined value of $p_i + q_i$. It is useful to note that the absolute value of eDKL, $|p_i \log_2 \frac{p_i}{q_i}|$, is asymmetric.
    \item The asymmetric patterns of eNewC and eNewGC are more intutive than that of eDJS since the right part of a curve (i.e., $\delta > 0$) is not expected to be lower than the left part (i.e., $\delta < 0$). For $q_a(z_a) = 0.5-\delta$ and $q_b(z_b) = 0.5+\delta$, the divergence at letter $z_b$ is expected to be no less than that at $z_a$ because the combined probability at $z_b$ is higher.
    \item the curve shapes of eDJS, eNewG, eNewGC correspond to that of eDKL better than eNew and eNewG.
\end{enumerate}

\section{Conclusions}
\label{sec:Conclusions}
In conclusion, given an alphabet $\mathbb{Z}$ with a finite number of letters and with a true PMF $P$ and an estimated PMF $Q$, the measure of their divergence should be bounded if one interprets the divergence based on the inefficiency caused by using $Q$ instead of $P$ in coding $\mathbb{Z}$.

This confirms that the unbounded term $\mathcal{D}_{\text{KL}}(\mathbb{Z}'_i || \mathbb{Z}_i)$ in Equation (\ref{eq:CBR}) should ideally be replaced with a bounded term. Semantically, the upper bound of the potential distortion should ideally be the maximum entropy of $\mathbb{Z}_i$, i.e., $\mathcal{H}_{\text{max}}(\mathbb{Z}_i)$. The lower bound should ideally be 0.
As the JS-divergence and different versions of the new divergence in Equations (\ref{eq:New}, \ref{eq:NewG}, and \ref{eq:NewGC}) are all bounded by [0, 1], they all meet the boundedness requirement with a scaling factor $\mathcal{H}_{\text{max}}(\mathbb{Z}_i)$.

If one prefers to preserve the curvature of the KL-divergence to some extent (e.g., based on the curve 0.3DKL in Figure \ref{fig:NonZero}), the JS-divergence and the generalized versions of the new divergence in Equations (\ref{eq:NewG},\ref{eq:NewGC}) (with $k=2$) are suitable candidates.

On the other hand, there is no fundamental reason to preserve the curvature of the KL-divergence.
The close-to-linear patterns shown by the basic version of the new divergence (i.e., Equation (\ref{eq:New})) in Figures \ref{fig:NonZero} and \ref{fig:NearZero} suggest that it may be easy to estimate or interpret such a measure mentally. Hence the new divergence in Equations (\ref{eq:NewG},\ref{eq:NewGC}) (with $k=1$) may have some advantages. 

Meanwhile, if one prefers to preserve the non-commutative property of the KL-divergence to some extent, the non-commutative version of the new divergence in Equations (\ref{eq:NewG}) is a suitable candidate.
On the other hand, if one prefers a commutative divergence measure, the JS-divergence and the commutative version of the new divergence in Equation (\ref{eq:NewGC}) are suitable candidates.

In addition, for measuring the divergence of an individual letter $z_i \in \mathbb{Z}$, the KL-divergence, JS-divergence, and the new divergence all have the common informative term in the form of $x \log (y)$, where $x$ and $y$ are some quantities.
As discussed in conjunction with Figure \ref{fig:Element}, if it is sensible to assume that this term should correlate to $x$ positively when $y$ is fixed, the commutative version of the new divergences in Equation (\ref{eq:NewGC}) exhibits such a positive correlation, while the JS-divergence exhibits a negative correlation. 

We appreciate that there will be different preferences in different applications, and it may take many years for different preferences to converge.

\bibliographystyle{abbrv}
\nocite{*}
\bibliography{boundedness}

\end{document}